# Finding and proving the exact ground state of a generalized Ising model by convex optimization and MAX-SAT

Wenxuan Huang[1], Daniil A. Kitchaev[1], Stephen Dacek[1], Ziqin Rong[1], Alexander Urban[3], Shan Cao[1], Chuan Luo[2], and Gerbrand Ceder[1,3,4]

1, Department of Material Science and Engineering, Massachusetts Institute of Technology, MA, USA

2, Key Laboratory of High Confidence Software Technologies, Peking University, Beijing, China

3, Department of Materials Science and Engineering, UC Berkeley, Berkeley, CA, USA

4, Materials Science Division, Lawrence Berkeley National Laboratory, Berkeley, CA, USA

## Abstract

Lattice models, also known as generalized Ising models or cluster expansions, are widely used in many areas of science and are routinely applied to alloy thermodynamics, solid-solid phase transitions, magnetic and thermal properties of solids, and fluid mechanics, among others. However, the problem of finding the true global ground state of a lattice model, which is essential for all of the aforementioned applications, has remained unresolved, with only a limited number of results for highly simplified systems known. In this article, we present the first general algorithm to find the exact ground states of complex lattice models and to prove their global optimality, resolving this fundamental problem in condensed matter and materials theory. We transform the infinite-discrete-optimization problem into a pair of combinatorial optimization (MAX-SAT) and non-smooth convex optimization (MAX-MIN) problems, which provide upper and lower bounds on the ground state energy respectively. By systematically converging these bounds to each other, we find and prove the exact ground state of realistic Hamiltonians whose solutions are completely intractable via traditional methods. Considering that currently such Hamiltonians are solved using simulated annealing and genetic algorithms that are often unable to find the true global energy minimum, and never able to prove the optimality of their result, our work opens the door to resolving long-standing uncertainties in lattice models of physical phenomena.

## Introduction

Lattice models have wide applicability in science [1-10], and are commonly used in a wide range of applications, such as magnetism [11], alloy thermodynamics [12], fluid dynamics[13], phase transitions in oxides [14], and thermal conductivity [15]. A lattice model, also referred to as generalized Ising model [16] or cluster expansion [12], is the discrete representation of materials properties, e.g., formation energies, in terms of lattice sites and site interactions. In first-principles thermodynamics, lattice models take on a particularly important role as they appear naturally through a coarse graining of the partition function [17] of systems with substitutional degrees of freedom. As such, they are invaluable tools to predict the structure and phase diagrams of crystalline solids based on a limited set of ab-initio calculations [18-22]. In particular, the ground states of a lattice model determine the 0K phase diagram of the system. However, the procedure to find and prove the exact ground state of a lattice model, defined on an arbitrary lattice with any interaction range and number of species remains an unsolved problem, with only a limited number of special-case solutions known in the literature [23-29]. In general systems, an approximation of the ground state is typically obtained from Monte Carlo simulations, which by their stochastic nature can prove neither convergence nor optimality. Thus, in light of the wide applicability of the generalized Ising model, an efficient approach to finding and proving its true ground states would not only resolve long-standing uncertainties in the field and give significant insight into the behavior of lattice models, but would also facilitate their use in ab-initio thermodynamics.

In this paper, we present an efficient algorithm that, in most cases, finds the global ground state of an arbitrary lattice model in any dimension and of any complexity, and proves the optimality of the solution. We first introduce the formal structure of a general lattice model and the Hamiltonian used to the describe it. We proceed to derive a solution to this optimization problem by converging a periodicity-constrained upper bound and aperiodic lower bound on the total energy. For calculating the upper bound, we derive an equivalence between the optimization of the Hamiltonian under a fixed periodicity and MAX-SAT pseudo-Boolean optimization (PBO), allowing us to leverage existing highly optimized and mathematically rigorous programing tools. To obtain the lower bound on the ground state energy, we derive an entirely new approach based on a maximization of minimum-energy local configurations. Finally, we demonstrate the accuracy, robustness and efficiency of our approach using both an assortment of random Hamiltonians and an example of a realistic application in materials science.

## Notation and Background

A lattice model is a set of fixed sites on which objects (spins, atom of different types, atoms and vacancies ,etc.) are to be distributed. Its energy or Hamiltonian consists of coupling terms between pairs, triplets, and other groups of sites, which we refer to as "clusters". A formal definition of effective cluster interactions can be found in [12]. Before discussing the algorithmic details of our method in the following sections, it is essential to establish a precise mathematical definition of a general lattice model Hamiltonian and the task of determining its ground states. The ground state problem can formally be stated as follows: Given a set of effective cluster interactions (ECI's) $J \in \mathbf{R}^{\mathbf{C}}$, where $\mathbf{C}$ is the set of interacting clusters and $\mathbf{R}$ is the set of real numbers, what is the configuration $s : \mathbf{D} \to \{0,1\}$, where $\mathbf{D}$ is the domain of configuration space, such that the global Hamiltonian $H$ is minimized:

$$\min_{s} H = \min_{s} \lim_{N \to \infty} \frac{1}{(2N+1)^3} \sum_{(i,j,k) \in \{-N,\ldots,N\}^3} \sum_{\alpha \in \mathbf{C}} J_\alpha \prod_{(x,y,z,p,t) \in \alpha} s_{i+x,j+y,k+z,p,t} \quad (1)$$

In the Hamiltonian of Eq. (1), each $\alpha \in \mathbf{C}$ is an individual interacting cluster of sites. In turn, each site within $\alpha$ is defined by a tuple $(x,y,z,p,t)$, wherein $(x,y,z)$ is the index of the primitive cell containing the interacting site, $p$ denotes the index of the sub-site to distinguish between multiple sub-lattices in that cell, and $t$ is the species occupying the site. To discretize the interactions, we introduce the "spin" variables $s_{x,y,z,p,t}$, where $s_{x,y,z,p,t} = 1$ indicates that the $p^{\text{th}}$ sub-site of the $(x,y,z)$ primitive cell is occupied by species $t$, and otherwise $s_{x,y,z,p,t} = 0$. The energy can be represented in terms of spin products, where each cluster $\alpha$ is associated with an ECI $J_\alpha$ denoting the energy associated with this particular cluster. To obtain the energy of the entire system, each cluster needs to be translated over all possible periodic images of the primitive cell, i.e., we have to consider all possible translations of the interacting cluster $\alpha$, defined as a set of $(x,y,z,p,t)$, by $(i,j,k)$ lattice primitive cells translations, yielding the spin product $\prod_{(x,y,z,p,t) \in \alpha} s_{i+x,j+y,k+z,p,t}$. Finally, the prefactor $\frac{1}{(2N+1)^3}$ normalizes the energy to one lattice primitive cell, and the limit of $N$ approaching infinity emphasizes our objective of minimizing the average energy over the entire infinitely large lattice. One remaining detail is that the Hamiltonian given in Eq. (1) is constrained such that that each site in the lattice must be occupied. For the sake of simplicity, lattice vacancies are included as explicit species in the Hamiltonian, so that all spin variables associated with the same site sum up to

one:

$$\sum_{t \in \mathbf{c}(p)} s_{x,y,z,p,t} = 1 \; \forall (x,y,z,p) \in \mathbf{F} \tag{2}$$

In Eq. (2), $\mathbf{F}$ is the set of all sites in the form of $(x,y,z,p)$, and $\mathbf{c}(p)$ denotes the set of species that can occupy sub-site $p$. The domain of configuration space $\mathbf{D}$ can be formally defined as the set of all $(x,y,z,p,t)$, with $t \in \mathbf{c}(p)$.

To further illustrate the notation introduced above, **Figure 1** depicts an example of a two-dimensional lattice Hamiltonian for a squared lattice with two sub-sites in each lattice primitive cell, i.e., $p \in \{0,1\}$. Each sub-site may be occupied by 3 types of species, so that $t \in \{0,1,2\}$, where $t=0$ shall be the reference (for example, vacancy) species. Hence, the energy of the system relative to the reference can be encoded into $t \in \{1,2\}$. Furthermore, the Hamiltonian shall be defined by only 2 different pairwise interaction types with the associated clusters $\alpha = \{(0,0,0,1,2),(1,2,0,0,1)\}$ and $\beta = \{(0,1,0,0,2),(0,0,0,1,2)\}$, and thus the set of all clusters is $\mathbf{C} = \{\alpha, \beta\}$. The first three of the five indices between "( )" brackets indicate the initial unit cell position, the forth index corresponds to the position in the unit cell (sub-site index), and the last index gives the species. The third component of the cell index (x,y,z) was retained for generality but set to 0 for this two-dimensional example. The example configuration shown in **Figure 1** depicts three specific interactions: The interaction represented on the bottom left in in the figure is of type $\alpha$ with $(i,j,k) = (0,0,0)$, corresponding to the spin product $J_\alpha s_{0,0,0,1,2} \cdot s_{1,2,0,0,1}$. The interaction in the center of the figure also belongs to type $\alpha$ but with $(i,j,k)=(1,1,0)$, corresponding to the spin product $J_\alpha s_{0+1,0+1,0,1,2} \cdot s_{1+1,2+1,0,0,1} = J_\alpha s_{1,1,0,1,2} \cdot s_{2,3,0,0,1}$. Lastly, the interaction on the right represents an interacting $\beta$ cluster, with $(i,j,k)=(3,0,0)$, yielding a spin product of $J_\beta s_{0+3,1,0,0,2} s_{0+3,0,0,1,2} = J_\beta s_{3,1,0,0,2} s_{3,0,0,1,2}$.

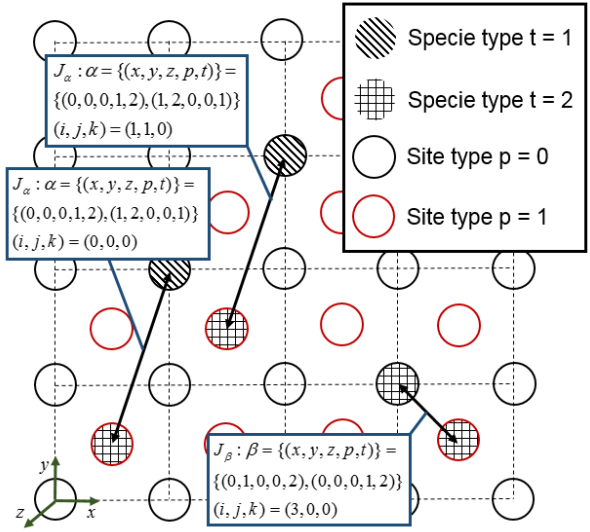

**Figure 1:** Illustration of a lattice Hamiltonian and exemplary cluster interactions. The primitive unit of the lattice is indicated by a thin dashed line, and sites are represented by circles. Two different site types are distinguished by black and red borders, respectively. The non-vacancy species that can occupy the sites are indicated by two different hatchings.

Currently, the most common approach to find the ground state of a generalized Ising model is simulated annealing [30] based on Metropolis Monte Carlo [31] in an *ad hoc* finite lattice cell. This approach has two major drawbacks. First, it is inherently an optimization over a finite set of sites, whereas the true objective function is defined over an infinite number of sites. Second, the result obtained from a Monte Carlo calculation is simply a particular low-energy configuration, a local minimum of energy, with no guarantee that it is the true ground state. This limitation becomes especially problematic when the size of the ground state structure increases, since the large number of degrees of freedom quickly renders it infeasible to sample the low-energy configurations in the cell. Hence, due to its stochastic nature and dependence on a particular lattice cell, simulated annealing can only identify possible ground state candidates, but it can never guarantee that the global ground state has been found.

An alternative approach that provides a provable ground state is the configurational polytope method [32,33] combined with vertex enumeration [34]. This method provides a beautiful reformulation of the ground state problem as linear programming. Unfortunately, this approach yields a polytope with an enormous number of "unconstructible" vertices, i.e. solutions that do not correspond to realizable lattice configurations [35,36], and there is no general, tractable algorithm to extract the true, constructible polytope. Recently, the "basic ray" method has been proposed, and used to obtain the ground states of several small systems [23-25]. However, a universal algorithm based on this method is not known [24], and the number of systems solved by this approach is limited.

In the following, we present a universal algorithm for finding and proving the exact ground state of a generalized Ising model. We derive the algorithm, demonstrate its universal applicability to arbitrary lattices and general multicomponent systems, and demonstrate its computational performance for practically relevant systems.

## Method formulation

Our general scheme for finding an exact ground state of an Ising model is to calculate and converge upper and lower bounds on the energy. We note that the energy of any periodic configuration is an upper bound on the ground state energy [37]. Thus, by enumerating periodicities and finding the exact ground state for each we can successively tighten the upper bound on the true ground state energy. If an exact periodic ground state structure exists, we are guaranteed to obtain the tightest upper bound possible once we reach the true periodicity by enumeration. However, there is no way of knowing when this condition has been reached, i.e., when the enumeration should stop. To be useful in practice, we therefore require an additional procedure to construct successfully tighter, rigorous lower bounds on the ground state energy, so that periodicity enumeration can be stopped when the upper and lower bounds match, indicating that the exact ground state has been found. Our approach for the construction of the lower bound is less intuitive and involves the optimization over a non-periodic domain. We discuss both upper and lower bound procedure separately.

**Obtaining the ground state at a fixed periodicity**

The first element of our solution to the ground state problem is to efficiently find the ground state given a fixed periodicity of the solution. While this problem is typically solved by Metropolis Monte Carlo (MC) simulated annealing in a prescribed simulation cell, this approach cannot prove that the periodic solution found is in fact optimal, not even for a given periodicity.

To arrive at a methodology that allows obtaining a provably optimal solution, we convert the problem of minimizing the Hamiltonian into a mathematical programming problem. The advantage of this approach is that mathematical programming algorithms not only yield good performance, but also require a rigorous proof of solution correctness, i.e. optimality, before termination. Classic examples of mathematical programming algorithms are the simplex method in linear programming [38] and the branch and bound method for integer programming [39], where the algorithm itself is also a schematic of the proof of optimality. As we will show in the following, the ground state problem for a fixed periodicity can be transformed into a maximum satisfiability problem [40], a well-researched class of optimization problems for which highly efficient solvers exist [41,42].

Using the notation introduced in the previous section, we note that calculating the periodic ground state is equivalent to solving the finite optimization problem:

$$\min_{\widehat{s}} \sum_{\alpha \in \widetilde{\mathbf{C}}} J_\alpha \prod_{(x,y,z,p,t) \in \alpha} \widehat{s}_{x,y,z,p,t} \quad (3)$$

subject to:

$$\sum_{t \in \mathbf{c}(p)} \widehat{s}_{x,y,z,p,t} = 1 \ \forall (x,y,z,p) \in \mathbf{F}_{\text{finite}} \quad (4)$$

where $\widehat{s}_{x,y,z,p,t}$ is, as $s$ in Eq. (2), the indicator variable of species $t$ on site $(x, y, z, p)$, with the difference that $\widehat{s}$ is now defined on a smaller domain determined by the periodicity, $\widetilde{\mathbf{C}}$ is the set of all interacting clusters within the fixed periodic system and $\mathbf{F}_{\text{finite}}$ is the set of sites within the fixed periodic unit cell. Such an optimization over discrete $\{0,1\}$ variables can be equivalently posed as a logic problem by converting the minimization problem into the negative of a maximization problem and replacing the discrete variables by Boolean equivalents. Following this insight, the minimization of the finite Hamiltonian can thus be expressed in the form of a pseudo-boolean optimization (PBO) problem, allowing us to solve this optimization as a weighted partial maximum satisfiability (MAX-SAT) [41,42] problem. The essence of MAX-SAT is to model the discrete optimization problem by maximizing the number of logical clauses that can be satisfied in a Boolean formula of conjunctive normal form, weighted by a set of arbitrary coefficients.

To illustrate this approach, we consider the example of a binary 1D system with a positive point term $J_0$ and a negative nearest-neighbor interaction $J_{NN}$, on a 2-site unit cell. For this system, the transformation is:

$$\begin{aligned}
E &= \min_{\widehat{s}_0,\widehat{s}_1}\left(J_0\widehat{s}_0 + J_0\widehat{s}_1 + J_{NN}\widehat{s}_0\widehat{s}_1\right)\\
&= -\max\left(J_0(1-\widehat{s}_0) - J_0 + J_0(1-\widehat{s}_1) - J_0 - J_{NN}\left(\widehat{s}_0\widehat{s}_1\right)\right)\\
&= -\max\left(J_0(\neg\widehat{s}_0) - J_0 + J_0(\neg\widehat{s}_1) - J_0 + (-J_{NN})\left((1-\neg\widehat{s}_0)\widehat{s}_1\right)\right)\\
&= -\max\left(J_0(\neg\widehat{s}_0) - J_0 + J_0(\neg\widehat{s}_1) - J_0 + (-J_{NN})\widehat{s}_1 + (-J_{NN})(1-\neg\widehat{s}_0\widehat{s}_1) - (-J_{NN})\right) \quad (5)\\
&= (2J_0 - J_{NN}) - \mathrm{MAXSAT}\left(J_0(\neg\widehat{s}_0)\wedge J_0(\neg\widehat{s}_1)\wedge(-J_{NN})(\widehat{s}_1)\wedge(-J_{NN})(\widehat{s}_0\vee\neg\widehat{s}_1)\right)
\end{aligned}$$

where the indicator variable $\widehat{s}_i$ is now also a Boolean variable in the MAX-SAT setting, and the $\wedge$, $\vee$ and $\neg$ operators correspond to logical "and", "or" and "not" respectively. Note that, although in a MAX-SAT problem the coefficient of each clause needs to be positive, it is still possible to transform an arbitrary set of cluster interactions $J_i$ into a proper MAX-SAT input, as in the example above.

The advantage of formulating the ground-state problem in this form is that MAX-SAT is one of the most actively researched NP-hard problems [43], allowing us to leverage the extensive literature written on the topic [44-47]. Note that any complete MAX-SAT solver encodes a proof of optimality [45] and includes a published proof of algorithm correctness (that it is guaranteed to find the optimal solution) [41] and efficiency [46,47]. Furthermore, the algorithms are run through an annual MAX-SAT competition [44], which tests their correctness, robustness and efficiency. Under such stringent criteria, by converting our problem into MAX-SAT, we can safely guarantee provability. By converting the ground state search problem into MAX-SAT, we are also able to further investigate the advanced proof schemes and fast algorithms developed over the last twenty years of MAX-SAT research [41,42,45]. The particular MAX-SAT solver we choose, based on the results of the MAX-SAT 2014 competition benchmarking and our own testing, is CCLS_to_akmaxsat [42,48].

Another notable advantage of MAX-SAT over MC for the solution of the ground-state problem is that state-of–the-art MAX-SAT solvers generally include sophisticated methods to escape from local minima [46,47] to arrive at the global minimum faster and more robustly than MC.

To verify the efficiency, robustness, and accuracy of the MAX-SAT solver compared to conventional MC, we performed a series of performance tests for both algorithms. We constructed a set of random 1D and 2D pair-interaction Hamiltonians with interactions up to the 28$^{th}$ nearest neighbor in 1D and up to the 10$^{th}$ nearest neighbor in 2D. We then attempted to find the ground state of each system by MAX-SAT and Monte Carlo. Finally, we considered only those Hamiltonians which could be classified as "difficult", which we define as having a ground state unit cell with more than 4 sites in 1D, or more than 12 sites in 2D. Among these "difficult" Hamiltonians, we find that MC is unable to find the ground state energy comparable to the MAX-SAT result in 10% of cases.

**Lower bound calculation**

The second element of our algorithm is the optimization of a lower bound to the ground state energy. The lower bound optimization provides both a proof of optimality of the ground state energy independent of periodicity, and a termination condition for the periodicity enumeration discussed in the previous section. To start, we prove that minimization of the Hamiltonian on a finite group of sites, without any periodic constraints, provides a lower bound for the ground state energy. To see why this statement is true, consider the bounds on the Hamiltonian:

$$H(s) = \lim_{N \to \infty} \frac{1}{(2N+1)^3} \sum_{\substack{(i,j,k) \in \\ \{-N,\ldots,N\}^3}} \sum_{\alpha \in \mathbf{C}} J_\alpha \prod_{(x,y,z,p,t) \in \alpha} s_{i+x,j+y,k+z,p,t} \quad \text{(6-a)}$$

$$\equiv \lim_{N \to \infty} \frac{1}{(2N+1)^3} \sum_{\substack{(i,j,k) \in \\ \{-N,\ldots,N\}^3}} E_{i,j,k,s} \geq \min_{i,j,k} E_{i,j,k,s} \geq \min_{\mathbf{s_0} \in \{0,1\}^\mathbf{B}} E_{\mathbf{s_0}} \quad \text{(6-b)}$$

where $E_{i,j,k,s}$ is defined as $\sum_{\alpha \in \mathbf{C}} J_\alpha \prod_{(x,y,z,p,t) \in \alpha} s_{i+x,j+y,k+z,p,t}$ and represents the energy of a block configuration in the lattice at location $(i,j,k)$ for a specific $s$. $\mathbf{B}$ is the block cluster containing the relevant $(x,y,z,p,t)$; formally $\mathbf{B} = \bigcup_{\alpha \in \mathbf{C}} \alpha$. Then, $\mathbf{s_0} \in \{0,1\}^\mathbf{B}$ is naturally defined as a block configuration and $E_{\mathbf{s_0}}$ as the energy corresponding to block configuration $\mathbf{s_0}$.

The first part of equation (6-b) is a restatement of the total average energy as an average over $(i,j,k)$ of block configuration energies $E_{i,j,k,s}$. As the LHS of equation (6-b) is an average over $(i,j,k)$ of $E_{i,j,k,s}$ it must be greater than or equal to the minimum over $(i,j,k)$ of $E_{i,j,k,s}$, the second part of equation (6-b). Hence, a minimization of $E_{i,j,k,s}$ over configuration space (the RHS), provides a lower bound:

$$H(s) \geq \min_{\mathbf{s_0} \in \{0,1\}^\mathbf{B}} E_{\mathbf{s_0}} \quad \text{(7)}$$

As an example, consider a simple binary 1D lattice system with interactions up to the next nearest neighbor (NNN). The energy of this system is bounded from below by the energy of the lowest energy block configuration:

$$H = \lim_{N \to \infty} \frac{1}{(2N+1)} \sum_{(i) \in \{-N,...,N\}^3} \left( J_0 s_i + J_1 s_i s_{i+1} + J_2 s_i s_{i+2} \right)$$
$$\geq \min_{s_0, s_1, s_2} \left( J_0 s_0 + J_1 s_0 s_1 + J_2 s_0 s_2 \right) \tag{8}$$

Thus, minimization over the block configurations $(s_0, s_1, s_2)$ produces a valid lower bound, $\min_{s_0, s_1, s_2} \left( J_0 s_0 + J_1 s_0 s_1 + J_2 s_0 s_2 \right)$, to the exact ground state energy.

Expressing the Hamiltonian in the above form assigns all weights of the point term interaction $J_0$ to the 0th site of the block cluster, corresponding to a $J_0 s_0$ term in the energy. We could have just as well redistributed the point term energy over all sites in the block cluster, transforming $J_0 s_0$ into $\frac{1}{3}(J_0 s_0 + J_0 s_1 + J_0 s_2)$, and similarly $J_1 s_0 s_2$ into $\frac{1}{2} J_1 (s_0 s_2 + s_1 s_3)$:

$$H = \lim_{N \to \infty} \frac{1}{(2N+1)} \sum_{(i) \in \{-N,...,N\}^3} \left( \begin{array}{c} \frac{1}{3}\left( J_0 s_i + J_0 s_{i+1} + J_0 s_{i+2} \right) \\ + \frac{1}{2} J_1 \left( s_i s_{i+1} + s_{i+1} s_{i+2} \right) + J_2 s_i s_{i+2} \end{array} \right)$$
$$\geq \min_{s_0, s_1, s_2} \left( \frac{1}{3}\left( J_0 s_0 + J_0 s_1 + J_0 s_2 \right) + \frac{1}{2} J_1 \left( s_0 s_2 + s_1 s_3 \right) + J_2 s_0 s_2 \right) \tag{9}$$

In the case of the exact infinite system Hamiltonian, this transformation simply corresponds to interchanging the order of the summation and thus imparts no difference to the total energy. However, in the case of the lower bound, we have obtained a new bounding condition, which is a key insight we will use to systematically obtain the tightest possible lower bound on the ground state energy.

**Tightening the lower bound using translationally equivalent ECIs**

Generally, the direct minimization of $E_{\mathbf{s}_0}$ as described in the previous section gives a very loose lower bound. In principle, a tighter lower bound could be generated systematically by enlarging the block size $|\mathbf{B}|$ used for finite minimization without periodicity constraints, thereby guaranteeing convergence to the exact ground state as we will show below. Furthermore by enlarging the periodicity used for periodic minimization, the upper bound energy is also guaranteed to converge to the exact ground state. To see why this statement is true, consider a minimization over larger and larger block clusters and the resulting block configuration $\mathbf{s}_0$. We could then translate and duplicate the configuration $\mathbf{s}_0$ to arrive at a periodic configuration $\mathbf{s}$ over the entire lattice. The energy of $\mathbf{s}$, $E_\mathbf{s}$, will differ from the configuration energy of $\mathbf{s}_0$, $E_{\mathbf{s}_0}$, only at the block boundaries, and the difference will diminish as blocks become larger and larger. Therefore the difference between $E_\mathbf{s}$ and $E_{\mathbf{s}_0}$ are approaching 0, while $E_\mathbf{s}$ is an upper bound and $E_{\mathbf{s}_0}$ is a lower bound of the exact ground state energy, proving that the lower bound energy $E_{\mathbf{s}_0}$ converges to the exact ground state energy. We also note that when we perform periodic minimization with the same periodicity as $\mathbf{s}$, we arrive at an upper bound energy smaller than or equal to $E_\mathbf{s}$, while greater than the lower bound $E_{\mathbf{s}_0}$, which proves that the upper bound also converges to the exact ground state energy with increasing periodicity.

Although in the limit of infinite block size the lower bound converges to the exact ground state energy, this approach is not practical, since finite minimization is NP-hard with respect to the block size. In the following, we present a much more efficient algorithm preserving the convergence property.

Given the original set of cluster interactions $J \in \mathbf{R}^{\mathbf{C}}$, a tighter lower bound can be obtained by introducing the set of equivalent $J_\lambda \in \mathbf{R}^{\overline{\mathbf{C}}}$, which will be defined so as to leave the Hamiltonian of the infinite system (1) unchanged, but will modify the Hamiltonian on a finite block. The $J_\lambda \in \mathbf{R}^{\overline{\mathbf{C}}}$ are parameterized by $\lambda \in \mathbf{R}^n$, which we define as a shift parameter.

Note that although the $J_\lambda$ will be defined to be equivalent to $J$ in the sense that they leave the global Hamiltonian unchanged, finite minimization without periodicity constraints does not yield the same lower bound. Thus, we can maximize the lower bound energy over $\lambda$ to obtain the tightest lower bound on the ground state energy: $\max_\lambda \min_{\mathbf{s} \in \{0,1\}^{\mathbf{B}}} E_{\lambda,\mathbf{s}}$ (10)

One natural way to introduce equivalent $J_\lambda$ is by redistributing an ECI over sites in the block cluster: given a fixed block cluster $\mathbf{B}$ to minimize over, for each cluster $\alpha \in \mathbf{C}$ such that $J_\alpha \neq 0$, we construct a set $C_\alpha$ such that all elements $\beta \in C_\alpha$ are equivalent to cluster $\alpha$ with respect to translations of the infinite lattice, and $\beta \subseteq \mathbf{B}$. For each element $\beta$ in $C_\alpha$, we assign weights $\lambda_\beta$ such that $\sum_{\beta \in C_\alpha} \lambda_\beta = 1$, which relate the translationally equivalent ECIs $J_\lambda$ to the original ECIs, so that for all $\alpha \in \mathbf{C}$ and $\beta \in C_\alpha$, $J_{\lambda,\beta} = \lambda_\beta J_\alpha$.

For example, returning to the 1D example of a NNN binary system given in Eq. (8), the conversion is:

$$\begin{aligned} H &= \lim_{N \to \infty} \frac{1}{(2N+1)} \sum_{(i) \in \{-N,\ldots,N\}^3} \left( J_0 s_i + J_1 s_i s_{i+1} + J_2 s_i s_{i+2} \right) \\ &= \lim_{N \to \infty} \frac{1}{(2N+1)} \sum_{(i) \in \{-N,\ldots,N\}^3} \begin{pmatrix} J_0 \left( \lambda_1 s_i + \lambda_2 s_{i+1} + (1 - \lambda_1 - \lambda_2) s_{i+2} \right) \\ + J_1 \left( \lambda_3 s_i s_{i+1} + (1 - \lambda_3) s_{i+1} s_{i+2} \right) + J_2 s_i s_{i+2} \end{pmatrix} \\ &\geq \min_{s_0, s_1, s_2} \left( J_0 \left( \lambda_1 s_0 + \lambda_2 s_1 + (1 - \lambda_1 - \lambda_2) s_2 \right) + J_1 \left( \lambda_3 s_0 s_1 + (1 - \lambda_3) s_1 s_2 \right) + J_2 s_0 s_2 \right) \end{aligned} \quad (11)$$

where the last expression provides a lower bound on the ground-state energy, dependent on $\lambda$. The rationale behind the $\lambda$ transform is analogous to that seen in Eq. (9): we exploit the fact that we can evenly distribute cluster interactions across sites, leaving the system unchanged, but obtaining a different lower bound on the ground state energy. Note that we are not limited to partitioning point terms equally over all sites, i.e., we could assign a contribution of the point term energy to site 0 with weight $\lambda_1$, to site 1 with $\lambda_2$ and to site 2 with $1 - \lambda_1 - \lambda_2$. In this way, we can generally convert $J_0 s_0$ to $J_0 \left( \lambda_1 s_0 + \lambda_2 s_1 + (1 - \lambda_1 - \lambda_2) s_2 \right)$, and $J_1 s_0 s_2$ into $J_1 \left( \lambda_3 s_0 s_1 + (1 - \lambda_3) s_1 s_2 \right)$, arriving at the lower bound expression of Eq. (11).

From this algorithm, we arrive at $\min_{\mathbf{s}\in\{0,1\}^{\mathbf{B}}} E_{\mathbf{s},\lambda}$, which is a lower bound dependent on $\lambda$. Thus,

$$\max_{\lambda} \min_{\mathbf{s}\in\{0,1\}^{\mathbf{B}}} E_{\lambda,\mathbf{s}} \qquad (12)$$

provides the maximal lower bound in the space defined by $\mathbf{B}$ and $\lambda$.

Finally, we note that Eq. (12) is a convex optimization problem. If $\mathbf{s}$ is fixed, $E_{\lambda,\mathbf{s}}$ is a linear function with respect to $\lambda$. Then $f(\lambda) = \min_{\mathbf{s}\in\{0,1\}^{\mathbf{B}}} E_{\lambda,\mathbf{s}}$ is the minimum of a set of linear functions evaluated at $\lambda$. Thus, $f(\lambda)$ is a concave function and $\max_{\lambda} f(\lambda)$ is a maximization over a concave function, which is equivalent to a minimization over a convex function, and thus is a convex optimization problem [49]. Due to its piece-wise linear characteristic, this problem belongs to the class of non-smooth convex optimization problems, where the objective function value is provided by MAX-SAT. In our implementation we use the level method [50] as a subclass of the bundle method [51] to efficiently solve this optimization.

## Further refinement of the lower bound

The introduction of equivalent $J_\lambda$ allows us to use finite minimization without periodicity constraints to obtain an exact lower bound on the ground state energy. Even when (11) can not provide the exact lower bound for small $|\mathbf{B}|$, by enlarging $|\mathbf{B}|$ and naturally introducing a higher dimensional $\lambda$ space, an exact lower bound can usually be obtained. For example, in the same 1D NNN binary system $|\mathbf{B}|$ can be enlarged and $\lambda$ space can be expanded as:

$$H = \lim_{N \to \infty} \frac{1}{(2N+1)} \sum_{(i) \in \{-N,\ldots,N\}^3} (J_0 s_i + J_1 s_i s_{i+1} + J_2 s_i s_{i+2})$$

$$\geq \min_{s_0, s_1, s_2, s_3} \begin{pmatrix} J_0 \left(\lambda_1 s_0 + \lambda_2 s_1 + \lambda_3 s_2 + (1 - \lambda_1 - \lambda_2 - \lambda_3) s_3 \right) + \\ J_1 \left(\lambda_4 s_0 s_1 + \lambda_5 s_1 s_2 + (1 - \lambda_4 - \lambda_5) s_2 s_3 \right) + \\ J_2 \left(\lambda_6 s_0 s_2 + (1 - \lambda_6) s_1 s_3 \right) \end{pmatrix} \quad (13)$$

Another alternative to introduce a refined lower bound without increasing $|\mathbf{B}|$ is by enlarging $\lambda$ space through the introduction of new clusters. For example:

$$H = \lim_{N \to \infty} \frac{1}{(2N+1)} \sum_{(i) \in \{-N,\ldots,N\}^3} (J_0 s_i + J_1 s_i s_{i+1} + J_2 s_i s_{i+2})$$

$$\geq \min_{s_0, s_1, s_2, s_3} \begin{pmatrix} J_0 \left(\lambda_1 s_0 + \lambda_2 s_1 + \lambda_3 s_2 + (1 - \lambda_1 - \lambda_2 - \lambda_3) s_3 \right) \\ + J_1 \left(\lambda_4 s_0 s_1 + \lambda_5 s_1 s_2 + (1 - \lambda_4 - \lambda_5) s_2 s_3 \right) \\ + J_2 \left(\lambda_6 s_0 s_2 + (1 - \lambda_6) s_1 s_3 \right) + \lambda_7 s_0 s_1 s_2 - \lambda_7 s_1 s_2 s_3 \end{pmatrix} \quad (14)$$

In cases where enlarging $|\mathbf{B}|$ is computationally very expensive, this second approach would be the only viable way to refine the lower bound. It remains unclear how such clusters should be introduced, given that there are, in general, an exponential number of them. However, this discussion is beyond the scope of the present article and will be addressed in future work. In this work, exact lower bounds are obtained by enlarging $|\mathbf{B}|$ as demonstrated in the first example.

We note that at the optimal $\lambda$ to the MAX-MIN optimization in (11), there are $N \cong \text{Dim}(\lambda)$ supporting hyperplanes at the optimal vertex. Thus, there are $N$ block configurations $s \in \{0, 1\}^\mathbf{B}$, which are optimal under such a $\lambda$-shift. If these $N$ block configurations can tile the whole space, the tiling is the exact ground state and the true ground state energy is equal to the lower bound value. Enlarging $\text{Dim}(\lambda)$ allows higher freedom in the tiling and thus provides a more accurate lower bound. As a result, in practice, exact lower bounds are usually obtainable without much computational expense by expanding $\text{Dim}(\lambda)$ naturally.

For aperiodic ground states whose representations are given in terms of tiling [52-54], this method provides a way to obtain the exact ground state, where periodicity enumeration (upper bound calculation) is not appropriate. Thus, this method could be useful in finding aperiodic ground states on a fixed lattice.

## Computational performance

To test the performance of this approach on practically relevant systems, we measure the runtime of our algorithm on binary 1D, 2D square, and 3D cubic lattices over random sets of ECIs across a spectrum of interaction ranges. First, we restrict ourselves to only pair interactions, calculating runtimes for up to 28 pair interactions on unit cells with up to 50 sites, where the energy of each interaction takes on a random value. In the 1D, 2D and 3D cases, this limit corresponds to all interactions up to and including the 28th, 10th and 5th nearest neighbors respectively. The results of these calculations on a single Intel E5-1650 3.20 GHz core are given in Figure 2. It is important to note that the code performance could be significantly improved by parallelization - the upper bound implementation is perfectly parallelizable up to at least hundreds of compute cores, and the lower-bound calculation parallelizes favorably based on the method chosen for the non-smooth convex optimization.

The results reveal that the primary source of runtime complexity is the range and number of interactions included in the Hamiltonian, with a secondary dependence on the dimensionality of the problem. As could be expected, increasing the range of interactions results in an exponential increase in runtime due to the exponential increase in the size of the spin configuration space. Fortunately, the increase in runtime with the number of interactions at a given range is polynomial. The effect of dimensionality is more subtle: dimensionality determines the number of distinct interactions at a given interaction range, and the number of possible unit cells containing no more than a set number of sites. We find that the former condition is important to the lower bound calculation runtime, while the latter condition determines the variation in the upper bound runtime.

In all cases, our implementation gives a very promising single-core runtime on the order of hours for realistic Hamiltonians, which typically include fewer than 100 interactions. The runtime scales more favorably when all the interactions included in the Hamiltonian are kept below some maximum range – for example, a Hamiltonian with 100 interactions limited to the $8^{th}$ nearest neighbor-range in 3D can be solved in 3 hours on a single core. This performance is consistent with the trends presented in Figure 2. In a 3D cubic system, there are 61 pair interactions at or below the 8th nearest neighbor range, which, based on the trend in Figure 2 would indicate a runtime of approximately $10^4$ seconds, or 2.7 hours. Thus, if we include three- and four- body terms in the Hamiltonian, the runtime is comparable to that of a pair-interaction Hamiltonian with the same interaction range.

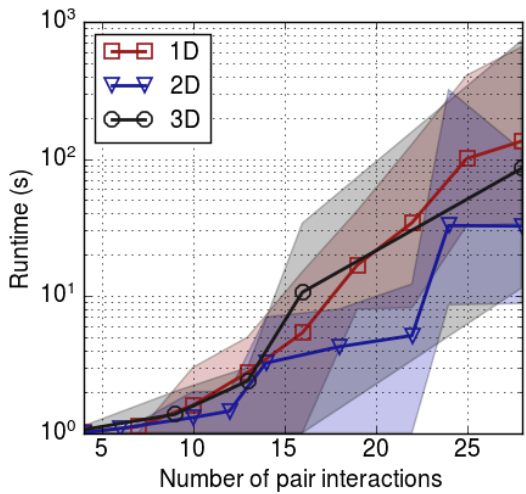

**Figure 2: Single-core computation time needed to find and prove the ground state of a 1D, 2D, and 3D pair-interaction Hamiltonian for unit cells up to 50 sites in size across an increasing range of pair-interactions. In all cases, the solver finds the ground state for all unit cells up to 50 atoms in size, and calculates a tight lower bound on the true ground state energy without enlarging |B|. Each point corresponds to the geometric average runtime of 100 such calculations with random interaction coefficients, while the shading gives the spread between the 20th and 80th percentiles.**

Finally, we apply our method to obtain the exact ground state of a cluster expansion Hamiltonian used to model sodium-vacancy orderings in the layered $Na_xNiO_2$ compound as a function of composition. The $J$ interactions for this system are determined from density-functional theory (DFT) calculations of 400 structures through standard approaches [12,55]. In this cluster expansion, there are 72 interacting clusters, including pair, triplet and quadruplet terms. We emphasize that no previous method exists that could in practice prove the exact ground states for a system with such interactions. In the configurational polytope method [32,33], a linear programming system with about $2^{32}$ variables and $2^{32}$ constraints would be required to capture the frustration effect necessary to provide a tight lower bound. Such a linear programming system cannot be solved in a practically relevant amount of time. In contrast, our method not only finds the exact ground states, but also proves their optimality on a time scale of minutes to hours.

As can be observed in Figure 3, our algorithm finds ground states at x=2/5, 1/2, and 3/5 that were not within the set of DFT input structures initially used to derive the cluster expansion. As we are able to prove that the solutions are optimal, we can guarantee that there are no other configurations of any unit cell size that are lower in energy. The inset shows the unusual ground state predicted at x=1/2 which is unlikely to be proposed from intuition.

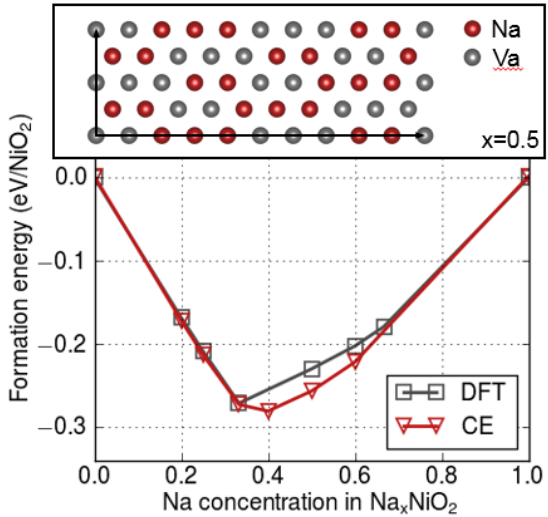

**Figure 3: Ground states found for a cluster expansion Hamiltonian of sodium-vacancy orderings in layered $Na_xNiO_2$. The red triangles indicate the mathematically proven ground states of the lattice model, whereas the gray squares are the originally proposed ground states from DFT calculations of 400 possible Na-vacancy arrangements. The ground state configuration for x=1/2 is shown in the inset.**

## Conclusion

We have introduced a MAX-MIN procedure to obtain the exact ground state of a generalized Ising model. To our knowledge, our approach is the only known method of proving the exact ground states of generalized Ising Hamiltonians with interactions of practically relevant complexity. In developing our procedure, we have derived an efficient approach to find an upper bound on the energy by transforming the finite optimization into a Boolean problem in the form of MAX-SAT, which provides a provable periodically-constrained optimum for the Hamiltonian. We then derived a lower bound on the energy from convex optimization over translationally-equivalent clusters. Finally, we converged the upper and lower bounds on the energy to find and prove the exact ground state. Mathematically, our approach exploits the tilability of minimum-energy local configurations to generate the exact global ground state. In practice, this procedure performs very well and has made it possible to determine the exact ground states of many formerly intractable systems, e.g. the cluster expansions of battery systems. We envision that in addition to resolving periodic ground states, the MAX-MIN procedure introduced here can serve as a tractable approach for resolving aperiodic ground states on a lattice.

## Acknowledgements

This work was supported primarily by the U.S. Department of Energy (DOE) under Contract No. DE-FG02-96ER45571. Additionally, the authors thank Prof. Stephen Boyd, Prof. Asu Ozdaglar, and Prof. Juan Pablo Vielma for inspiration on convex optimization. We are grateful to Prof. Joseph O'Rourke and Prof. Joel David Hamkins for guidance on tiling and complexity theory. Further, we thank Professor Carlos Ansótegui, Dr. Josep Argelich, and Dr. Adrian Kuegel for kindly providing us different MAXSAT solvers.